\documentclass[prd,aps,11pt,preprintnumbers,amssymb,amsmath,amsfonts,nofootinbib]{revtex4-2}
\usepackage{latexsym}
\usepackage{ulem}
\usepackage[utf8]{inputenc}
\usepackage{cancel}
\newcommand{\ba}{\begin{eqnarray}}
\newcommand{\ea}{\end{eqnarray}}

\newcommand{\be}{\begin{equation}}
\newcommand{\ee}{\end{equation}}

\usepackage[x11names,dvipsnames,hyperref]{xcolor}

\usepackage{slashed}
\usepackage{pifont}
\usepackage{dcolumn}
\usepackage{graphics}                                                                         
\usepackage[dvips]{graphicx}
\usepackage{multirow}
\usepackage{makecell}
\usepackage{bm}

\begin{document}
                                                                  
\date{\today}
\title{A two-body femtoscopy approach to \\ the proton-deuteron correlation function}

\author{Juan M. Torres-Rincon$^1$, \`Angels Ramos$^1$, Joel Rufí$^2$ }

 \affiliation{$^1$Departament de F\'isica Qu\`antica i Astrof\'isica, Facultat de F\'isica,  Universitat de Barcelona, Mart\'i i Franqu\`es 1, 08028 Barcelona, Spain and Institut de Ci\`encies del Cosmos (ICCUB), Facultat de F\'isica,  Universitat de Barcelona, Mart\'i i Franqu\`es 1, 08028 Barcelona, Spain}
 \affiliation{$^2$ Facultat de F\'isica,  Universitat de Barcelona, Mart\'i i Franqu\`es 1, 08028 Barcelona, Spain }
 
\date{\today}

\begin{abstract}
The proton-deuteron correlation function measured by the ALICE collaboration in high multiplicity p+p collisions shows a momentum dependence which is in contradiction with the predictions of the Lednick{\'y}-Lyuboshitz formalism of the two-body interaction. This result motivated a more sophisticated three-body description in terms of a composite deuteron. Encouraged by the good description of other deuteron observables in the context of heavy-ion collisions, we revisit this correlation function under the two-body approximation without relying on the Lednick{\'y}-Lyuboshitz approximation, but using the solution of the Schr\"odinger equation by incorporating both strong and Coulomb interactions. The two-body description provides a reasonable agreement to ALICE data for natural values of the source size. Interpreting the previously assumed attractive character of the $(L,S)=(0,3/2)$ channel of the $pd$ interaction into a more-likely repulsion, the results agree better not only with the ALICE measurements but also with recent STAR data in Au+Au collisions.
\end{abstract}

\maketitle


\section{Introduction}
\label{sec:intro}

Femtoscopy has emerged as a powerful technique in relativistic heavy-ion collisions (RHICs) to gain information of the system spacetime properties as well as details on hadronic interactions~\cite{Koonin:1977fh,Bauer:1992ffu,Baym:1997ce,Wiedemann:1999qn,Heinz:1999rw,Lisa:2005dd}. For a recent review on this field see Ref.~\cite{Fabbietti:2020bfg}. Accessing the details of particle interactions is most appealing for systems in which scattering experiments cannot be performed. Typical examples are particles that decay fast due to weak interactions and stable beams cannot be conceived. Femtoscopy of hyperons and other heavy particles with strangeness or charm have already been analyzed in the HADES experiment at GSI~\cite{HADES:2016dyd}, STAR experiment at BNL~\cite{STAR:2014dcy,STAR:2018uho}, and ALICE experiment at LHC~\cite{ALICE:2017jto,ALICE:2017iga,ALICE:2018nnl,ALICE:2021ovd,ALICE:2020wvi,ALICE:2019gcn,ALICE:2018ysd,ALICE:2021njx,ALICE:2019buq,ALICE:2018ysd,ALICE:2019eol,ALICE:2019hdt,ALICE:2020mfd,ALICE:2020mkb,ALICE:2021cpv,ALICE:2019igo,ALICE:2021cyj,ALICE:2024bhk}. 

Femtoscopy of nuclei has also become possible in which deuteron or more massive particles have been used. Recently, femtoscopy studies involving proton and deuteron pairs have been carried out in high-multiplicity p+p collisions at $\sqrt{s}=13$ TeV by the ALICE~\cite{ALICE:2023bny} experiment, in Au+Au collisions at $\sqrt{s_{NN}}=3$ GeV measured by the STAR experiment~\cite{STAR:2024lzt}, and in Ag+Ag collisions at $\sqrt{s_{NN}}=2.55$ GeV by the HADES experiment~\cite{Stefaniak:2024eux}.

In the case of p+p collisions produced at the LHC~\cite{ALICE:2023bny}, the measured $p-d$ correlation function has been compared by the ALICE Collaboration with the theoretical prediction based on the so-called Lednick{\'y}-Lyuboshitz (LL) approximation~\cite{Lednicky:1981su,Lednicky:2005tb}, which uses the two-body $L=0$ wave function in its asymptotic form by taking the low-energy scattering parameters (scattering length and scattering range) as the only inputs. These parameters were fixed from the experimental scattering data of Ref.~\cite{Arvieux:1974fma}. We review this approximation in App.~\ref{app:LL}. As seen in~\cite{ALICE:2023bny}, this approximation completely fails in describing the experimental correlation function with totally opposed results. In the LL approximation, the strong interaction dominates over the Coulomb barrier due to the large values of the scattering length (specially in the very attractive $^4S_{3/2}$ channel). On the contrary, the experimental data show little modifications of the repulsive electromagnetic interaction due to the strong interaction. 

It was advocated in Ref.~\cite{Viviani:2023kxw} that a full three-body description of the $p-d$ scattering problem was needed~\cite{Mrowczynski:2019yrr}. This would account for the intrinsic structure of the deuteron, and a more complex formalism was developed for this goal. The final result---also incorporating higher partial waves than $L=0$---shows a very good agreement with data~\cite{ALICE:2023bny}, arriving at the conclusion that accounting for the internal structure of the deuteron is essential to describe the $pd$ system.

While the size of the deuteron is comparable to the typical range of the emitting source, the use of a point like deuteron in RHIC studies has given results that are consistent with experimental data on several observables~\cite{Danielewicz:1991dh}: yields and spectra~\cite{Oh:2009gx,Oliinychenko:2020znl}, elliptic flow~\cite{Oh:2009gx,Staudenmaier:2021lrg}, etc. Also the freeze-out deuteron multiplicity are successfully predicted by the statistical thermal model~\cite{Andronic:2010qu,Andronic:2017pug}, where the deuteron is modelled on equal footing as the other hadrons. For this reason one would expect that the low-energy description of the $pd$ system with a elementary deuteron should be still possible as a first approximation to the real physics. We believe that before discarding the two-body approach in favor of a three-body formalism (without diminishing the value of that approach), one needs to (in)validate the two-body description by addressing it beyond the LL approximation.  

In this paper, we revisit the proton-deuteron problem by solving the Schr\"odinger equation with strong and Coulomb interactions and compute the full wave function valid in full space, instead only in the asymptotic region. In the standard application of the LL approach, it only considers low-energy scattering parameters in the $L=0$ partial wave. In our current approach, we are not limited by the low-energy expansion of the scattering amplitude and include higher partial waves beyond the $S$ wave.

\section{Proton-deuteron interaction: \\ Woods-Saxon and Gaussian potentials}

We start by setting the two-body interaction potentials between a proton and a deuteron using experimental information from the phase shifts at low energies. We will use data in different spin ($S$) and angular momentum ($L$) channels from Ref.~\cite{Arvieux:1974fma}. 

To access the wave function in a given channel, we solve the quantum two-body problem using a potential in position space $V(r)$ by means of the Schr\"odinger equation,
\begin{equation}
    -\frac{\hbar^2}{2\mu} \nabla^2 \psi(\bm{r}) + V(r) \psi(\bm{r}) = E \psi(\bm{r}) \ , \label{eq:schoedinger}
\end{equation}
where $\mu$ is the reduced mass of the $pd$ system and $E= \hbar^2 k^2/(2\mu) $ is the total energy of the system in the center of mass frame. We analyze the solution of Eq.~(\ref{eq:schoedinger}) by looking for bound states ($E<0$) as well as for scattering states, which we eventually use in the correlation function calculation (see Sec.~\ref{sec:corre}).

We use two sets of phenomenological potentials. The first one is taken from Ref.~\cite{Jennings:1985km}, and it has a Woods-Saxon (WS) form, 
\begin{equation}
    V_{ \textrm{WS}}(r) =  \frac{V_0}{1 + {\rm e}^{(r - R)/a}} \ , \label{eq:WS}
\end{equation} 
where $V_0$, $a$, and $R$ are parameters that depend on $S$ and $L$. The total spin of the $pd$ pair can be $S=\{ 1/2,3/2 \}$, while we will take orbital angular momentum values up to $L=2$, which is the maximum partial wave analyzed in Ref.~\cite{Arvieux:1974fma}.

\begin{table}[ht!]
\centering
\vspace*{2mm}
\begin{tabular}{|c|c|c|c|c|c|c|}
\hline
&   \multicolumn{3}{c|}{$S=1/2$} & \multicolumn{3}{c|}{$S=3/2$}  \\
\hline
 & $L=0$ & $L=1$& $L=2$ & $L=0$ & $L=1$ & $L=2$ \\
\hline
$V_0$ (MeV) & $-29.754$ & $-8.214$ & $-7.849$ & $-18.115$ & $-13.10$ & $14.878$ \\
$R$ (fm) & $2.826$ & $2.962$ & $2.974$ & $2.837$ & $2.067$ & $2.527$ \\
$a$ (fm) & $1.187$ & $0.259$ & $0.991$ & $0.9655$ & $1.578$ & $1.235$ \\
\hline
\hline
$E_B$ (MeV) & -7.780 &  &  & -2.539 &  &  \\
\hline
\end{tabular}
\caption{Values of the potential depth ($V_0$), the size ($R$), and the surface thickness parameter ($a$) of the Woods-Saxon potential in Eq.~(\ref{eq:WS}) for the proton-deuteron interaction taken from Ref.~\cite{Jennings:1985km}. They are  fitted to reproduce the experimental $pd$ scattering phase shifts in Ref.~\cite{Arvieux:1974fma}. Bottom row: binding energies of the bound states found by solving the Schr\"odinger equation in the different $(S,L)$ channels.\label{tab:potential}}
\end{table}

The parameters in Eq.~(\ref{eq:WS}) are given in Ref.~\cite{Jennings:1985km}, and they were calculated from fits to the scattering data in Ref.~\cite{Arvieux:1974fma}. There is a total of 18 parameters (3 for each spin-orbit channel) which we reproduce for convenience in Table~\ref{tab:potential}. In this table we also quote the values of the binding energies that we obtain for the bound states---without incorporating Coulomb interaction---that appear in different channels. Apart from the state with $S=1/2$, which could be identified with the $^3$He nucleus~\cite{Tomio:1987zz}, there is another bound state with smaller binding energy in the $S=3/2$ channel, which cannot be associated to any known physical state as $^3$He has no known nuclear excitations.

In the case of states in the continuum, we turn on the Coulomb potential and solve the Schr\"odinger equation as a function of the relative momentum $k$,
\begin{equation}
    \left(\frac{d^2}{dr^2}+k^2 - U(r) - \frac{L(L+1)}{r^2}\right)u_{L}(r,k) = 0 \ ,
\end{equation}
where $U(r) = 2\mu [V(r)+\alpha/r]$, with $\alpha$ being the fine structure constant, and $L$ is the orbital angular momentum. We solve the equation for $u_L(r)$ using the Numerov algorithm~\cite{notesnumerov}, and the solution in the outer region of the potential is matched to the asymptotic form~\cite{Macedo-Lima:2023fzp,joachain1975quantum},
\begin{equation}
    u_L(r,k) = k^{-1} e^{i\Delta_L} \left[\cos{\hat{\delta}_L} F_L(\gamma;kr) - \sin{\hat{\delta}_L}G_L(\gamma;kr)\right] \ ,
    \label{eq:ucs}    
\end{equation} 
where $\gamma=\mu\alpha/k$, $F_L (\gamma;kr)$ and $G_L (\gamma;kr)$ are the regular and irregular Coulomb  wave functions, respectively. These are incorporated in our code using the subroutines of Ref.~\cite{Salvat}. Finally, $\Delta_L = \sigma_L + \hat{\delta}_L$ is a global phase shift that contains the Coulomb phase shift, $\sigma_L= \arg \Gamma(1+L +i\gamma)$, and an additional phase shift, $\hat{\delta}_L$, that incorporates the effect of the strong interaction (but not equal to the pure strong phase shift~\cite{joachain1975quantum}).

\begin{figure}[ht]
  \centering
  \includegraphics[width=0.9\textwidth]{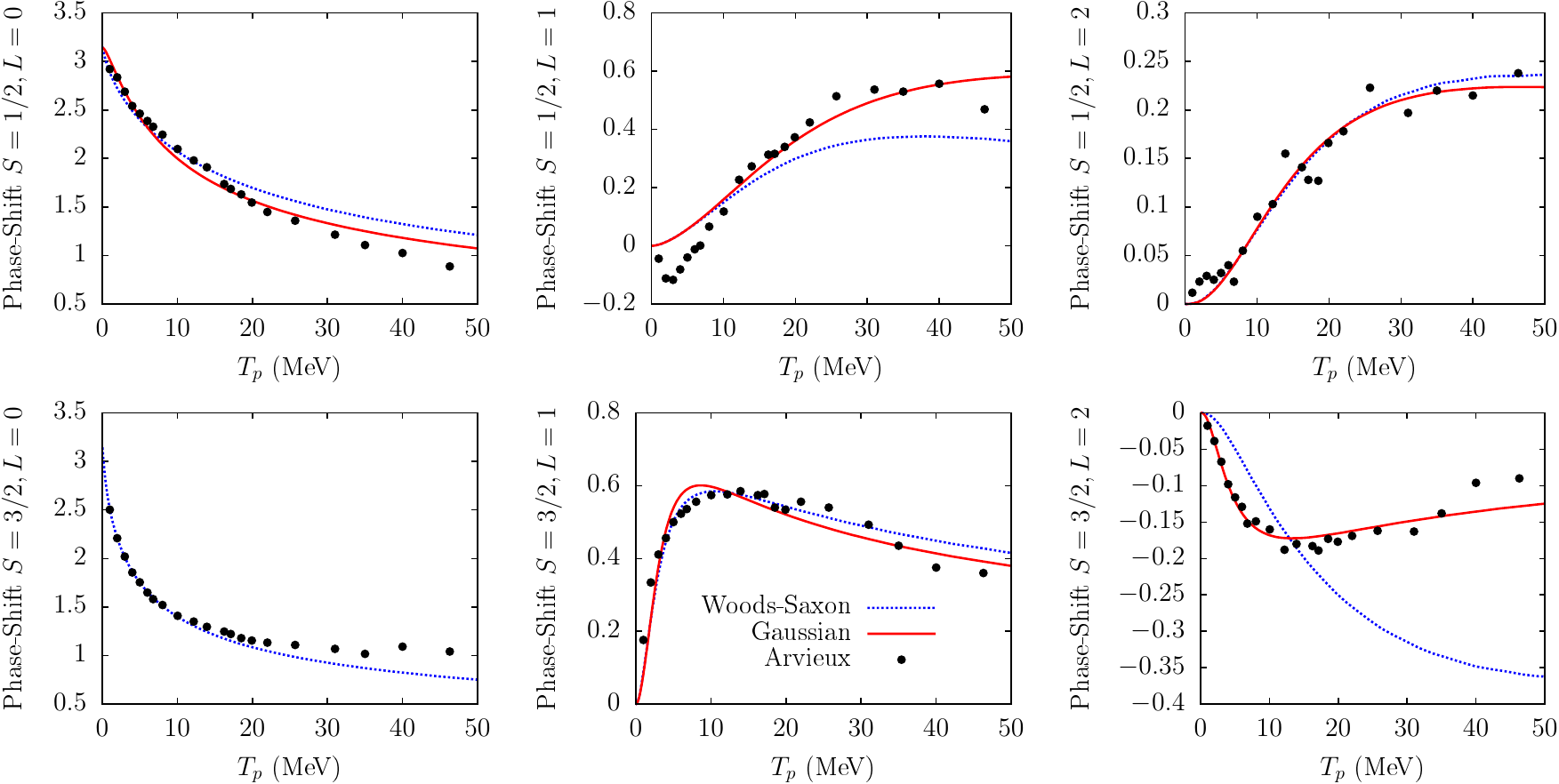}
  \caption{Results of the calculated $pd$ scattering phase shifts using the WS (dotted blue lines) and Gaussian (solid red lines) potentials, together with the experimental results from Ref.~\cite{Arvieux:1974fma}. (The Gaussian fit to the $S=3/2,L=0$ phase-shift is shown in Fig.~\ref{fig:phaseshifts_rep}.)}
  \label{fig:phaseshifts}
\end{figure}

The resulting phase shifts $\hat{\delta}_L$ (with Coulomb effects) using the WS potentials are compared in Fig.~\ref{fig:phaseshifts} (dotted blue lines) to the experimental data extracted from Ref.~\cite{Arvieux:1974fma} as functions of the kinetic energy of the incident proton in the laboratory frame, $T_p= E (m_p+m_d)/m_d$, with $m_p$ and $m_d$ being the proton and deuteron masses, respectively.

From Fig.~\ref{fig:phaseshifts} we observe a general good agreement of the results using the WS potential given in Ref.~\cite{Jennings:1985km} with the experimental phase shifts, except sizable deviations in the $S=1/2,L=1$ and $S=3/2,L=2$ cases. It is important to notice that in the $S=3/2,L=0$ channel the phase shift starts at $\pi$ and decreases, a signature of the bound state in that channel (see Table~\ref{tab:potential}). However, the  existence of such a state is dubious, and the attractive nature of this channel can be questioned due to the ambiguity of $\pi$ in the extraction of the scattering phase shifts. 

For this reason we try an alternative parametrization of the same experimental phase shifts but reinterpret the $S=3/2, L=0$ channel. On one hand we would like to use a simple form of the potentials with a lesser number of parameters than the WS parametrization. Therefore, we use a Gaussian form,
\be V_{\textrm{Gauss}}(r) = \bar{V}_0 \ \exp \left( - \frac{r^2}{r_G^2} \right) \ ,  \label{eq:Gauss} \ee
where the potential strength $\bar{V}_0$ and the potential range $r_G$ are parameters depending on $S$ and $L$.

On the other hand, we reanalyze the $S=3/2,L=0$ channel and consider it to be repulsive due to the absence of any physical bound state with $J=3/2$. Then we shift down the quoted $^4S_{3/2}$ phase shift in~\cite{Arvieux:1974fma} by $\pi$ in all its energy domain. This interpretation is consistent with the $S=3/2,L=0$ phase shift considered in other references~\cite{VanOers:1967oww,Eyre:1976dp,Huttel:1983wkj,Kievsky:1996ca} which starts at zero and decreases with energy, showing a pure repulsive interaction.

\begin{table}[ht!]
\centering
\vspace*{2mm}
\begin{tabular}{|c|c|c|c|c|c|c|}
\hline
&   \multicolumn{3}{c|}{$S=1/2$} & \multicolumn{3}{c|}{$S=3/2$}  \\
\hline
 & $L=0$ & $L=1$& $L=2$ & $L=0$ & $L=1$ & $L=2$ \\
\hline
$\bar{V}_0$ (MeV) & $-15.7 $ & $-49.0 $ & $-7.4 $ & $28.8 $ & $-6.8 $ & $1.4 $ \\
$r_G$ (fm) & $5.3 \pm 0.3$ & $1.7 \pm 0.3$ & $3.5 \pm 0.2$ & $3.9 \pm 0.2$ & $4.5 \pm 0.3$ & $7.4 \pm 1.2$ \\
\hline
\hline
$E_B$ (MeV) & -7.87 &  &  &  &  &  \\
\hline
\end{tabular}
\caption{Parameters of the $pd$ Gaussian potential of Eq.~\eqref{eq:Gauss} obtained by fitting experimental phase shifts of Ref.~\cite{Arvieux:1974fma} for the considered $(S,L)$ channels. Bottom row: Binding energy of the bound state found in the $S=1/2,L=0$ channel.}\label{tab:gaussian}
\end{table}

We perform fits of the experimental phase shifts in Ref.~\cite{Arvieux:1974fma} using the potential in Eq.~(\ref{eq:Gauss}) for each channel using the MINUIT package to generate the best fit parameters. The resulting fitting parameters of the Gaussian potentials are shown in Table~\ref{tab:gaussian}. In this table, we also provide the binding energy of the obtained bound state in the $S=1/2, L=0$ channels (without the Coulomb force). We find the state identified with the $^3$He to have a similar binding energy compared to that in the WS case. As expected, we find no bound state in the repulsive channel $S=3/2,L=0$. 

The resulting phase shifts obtained in the fit of Gaussian potentials are shown in Fig.~\ref{fig:phaseshifts} in solid red lines. Using a lesser number of parameters (12 vs 18), we are able to obtain satisfactory fits and even improve the description of the $S=1/2,L=1$ and $S=3/2,L=2$ channels, as compared to the WS case.

\begin{figure}[ht]
  \centering
  \includegraphics[width=0.4\textwidth]{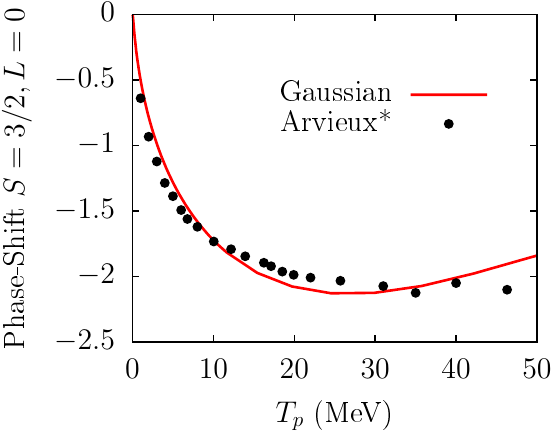}
  \caption{Phase shifts of the $pd$ scattering in the $S=3/2,L=0$ channel using the Gaussian potential in Eq.~(\ref{eq:Gauss}). Symbols are the experimental result of~\cite{Arvieux:1974fma} for this channel, but incorporating a down shift of $\pi$---due to the $\pi$-ambiguity of the phase-shift---to reinterpret the result as a repulsive interaction~\cite{VanOers:1967oww}.}
  \label{fig:phaseshifts_rep}
\end{figure}

The $S=3/2,L=0$ phase shifts calculated with the Gaussian potential are plotted in the separate Fig.~\ref{fig:phaseshifts_rep}. We superimpose the experimental data from Ref.~\cite{Arvieux:1974fma} shifted down a value of $\pi$ (denoted as Arvieux$^*$) to match the results with the repulsive interaction described in Ref.~\cite{VanOers:1967oww}. The fit is rather good and provides a repulsive potential which, as we said, admits no bound state as opposed to the WS case.

\section{Proton-deuteron correlation function~\label{sec:corre}}

Once we obtain the $pd$ wave functions for any value of $k$ as a function of $r$, we can address the femtoscopy correlation function. To compute it we use the Koonin-Pratt equation~\cite{Koonin:1977fh,Pratt:1990zq},
\begin{equation}
    C(k) = \int S( \bm{r}) \left| \Phi( \bm{r};\bm{k}) \right|^2 d^3r \ ,
    \label{corr}
\end{equation}
where $S( \bm{r})$ is the source function depending on the relative distance of the pair and $\Phi( \bm{r}; \bm{k})$ is the two-body wave function in the CM frame. For the source function, we choose a normalized Gaussian profile that depends only on the relative distance $r$~\cite{Fabbietti:2020bfg}, 
\begin{equation}
    S(\bm{r}) = \left(4\pi r_0^2\right)^{-3/2}\exp{\left(-\frac{r^2}{4r_0^2}\right)} \ , \label{eq:source}
\end{equation}
where $r_0$ is the radius parameter that defines the typical distance between a proton and a deuteron at freeze-out. In our calculations this quantity is estimated using experimental conditions (depending on the colliding system, collision centrality, etc.) or used as a free parameter. 

The total wave function includes both strong and Coulomb interactions, and it can be rewritten in the following convenient form by using a partial wave sum,
\begin{equation}
    \Phi(r,z;k) =  \Phi^{\textrm{C}}(r,z;k) + \sum_L (2L+1) \left[ \Phi_L(r;k) - \Phi_{L}^C(kr) \right]  P_L (\cos \theta)  \ , \label{eq:wavefull}
\end{equation}
where $z=r\cos\theta$, and $\theta$ the relative angle between $\bm{r}$ and $\bm{k}$. In Eq.~\eqref{eq:wavefull} the complete Coulomb wave function is given by~\cite{joachain1975quantum} (we use the notation used in Ref.~\cite{Torres-Rincon:2023qll}),
\begin{equation}
    \Phi^{\textrm{C}}(r,z;k)= e^{-\pi \gamma/2} \Gamma(1+ i\gamma) e^{ikz} \ {_1}F_{1}(-i\gamma;1;ik(r-z)) \ , \label{eq:completeCoulomb}
\end{equation}
with $\Gamma(a)$ being the Euler's gamma function and ${_1}F_{1}(a,b;c)$ the confluent hypergeometrical function or Kummers function. The $L$ partial-wave of the Coulomb wave function is given by~\cite{joachain1975quantum},
\begin{equation}
    \Phi_{L}^{\textrm{C}}(kr)= (kr)^{-1} i^L e^{i\sigma_L} F_L(\gamma;kr) \ .
\end{equation}

Inserting the square of (\ref{eq:wavefull}) into the correlation function, and using that the source is spherically symmetric, one gets from Eq.~(\ref{corr}),
\begin{equation}
    C(k) = \int S(r) \left|\Phi^{\textrm{C}}(r,z;k)\right|^2 d^3r  + \int 4\pi r^2 S(r) \sum_{L=0}^{L_{\textrm{max}}} (2L+1) \left[ \left|\frac{u_L(r,k)}{r}\right|^2 - \left|\Phi_{L}^{\textrm{C}}(kr)\right|^2\right] dr \ , 
\end{equation}
where $u_L (r,k)$ is the reduced radial wave function with angular momentum $L$, computed by the method detailed in the preceding section. Notice that we have truncated the partial wave sum up to a maximum value $L_{\textrm{max}}$ assuming that the sum converges.

We present the results for the correlation function by adding one by one the different partial waves, using a fixed value of the source radius, $r_0$, and performing the averaging over the $S=1/2$ and $S=3/2$ spins,
\begin{equation}
    C(k) = \frac{ {^2}C(k) + 2\, {^4}C(k) }{3} \ .
\end{equation}

\begin{figure}[ht]
  \centering
  \includegraphics[width=0.4\textwidth]{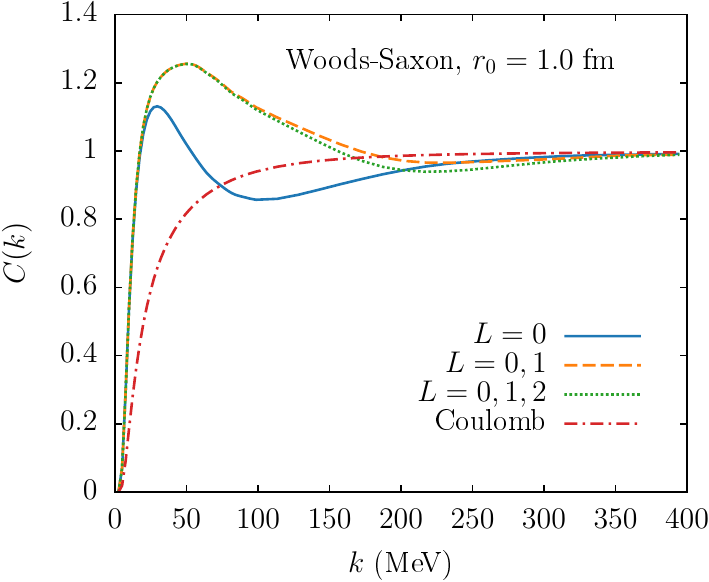}
  \includegraphics[width=0.4\textwidth]{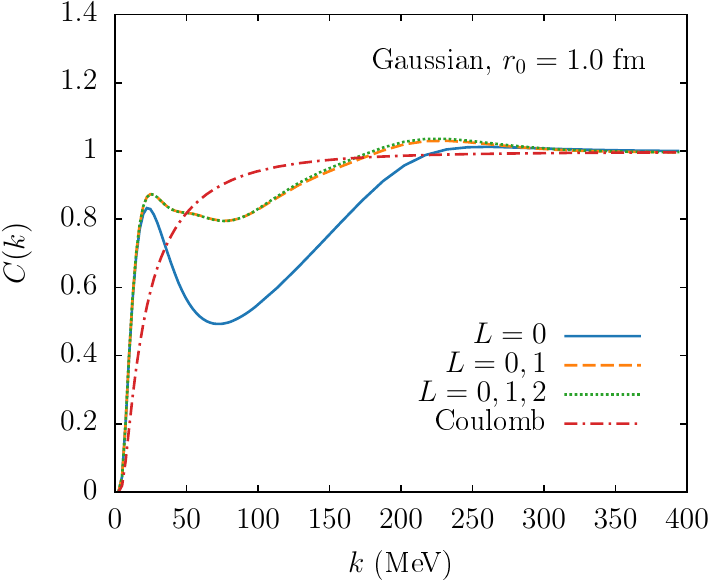}
  \includegraphics[width=0.4\textwidth]{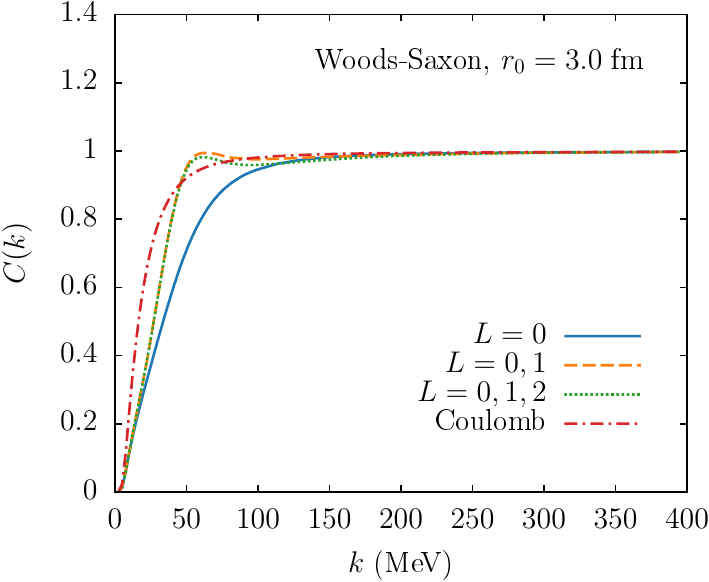}
  \includegraphics[width=0.4\textwidth]{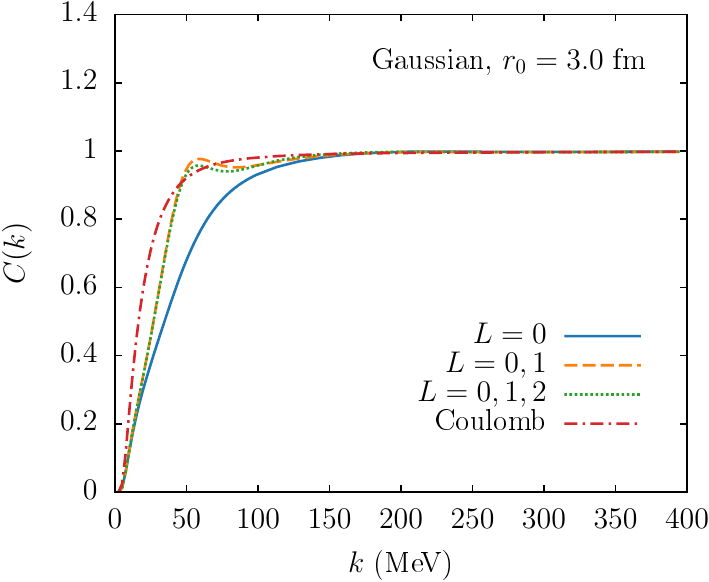}
  \caption{$pd$ correlation function for the WS (left column) and Gaussian potential (right column), obtained by adding successive partial waves $L=0,1,2$. Top panels: Source radius of $r_0 = 1$ fm. Bottom panels: $r_0=3$ fm.}
  \label{fig:Cqseparated}
\end{figure}

The results are shown in Fig.~\ref{fig:Cqseparated}. In the left column we present the results using the WS potentials and in the right column the results of the Gaussian potentials. For the top panels we use a small radius of $r_0=1.0$ fm typical from p+p collisions, while in the bottom panels we use $r_0=3$ fm, more consistent with mid peripheral heavy-ion collisions.

The dotted-dashed red line is the result of using the Coulomb interaction only. It serves as a baseline that reflects the electromagnetic repulsion of the pair, signalled by the correlation function lower than unity~\cite{Fabbietti:2020bfg}. The solid blue, dashed orange and dotted green lines are the results adding the strong interaction up to partial waves $L^{\textrm{max}}=0,1,2$, respectively. 

We observe that the main effect comes from the $L=0$ wave, which presents a bound state in the $S=1/2$ channel. In addition, the WS case also admits a bound state in the $S=3/2,L=0$ channel. The effect of adding the $L=1$ partial wave is important for both models, producing a visible increase around $k=100$ MeV. The effect of the $L=2$ partial wave is very small in both models and for both Gaussian radii, showing a fast convergence of the partial wave expansion.

We have seen that different potentials produce variations in the wave functions for the same $(S,L)$ close to $r=0$. In turn, this effect produces differences in the correlation function. With larger values of the source size $r_0$, the part of the wave function that contributes more to the correlation function is shifted toward the asymptotic region. Since the potentials are fitted with information of that precise region i.e. the phase shifts, then the resulting correlation functions are alike (except for the different phase shifts of the $S=3/2,L=0$ channel, as commented). In fact, the LL approximation should work better for larger radii, since the relevant part of the wave function becomes closer to the asymptotic part.

\begin{figure}[ht]
  \centering
    \includegraphics[width=0.45\textwidth]{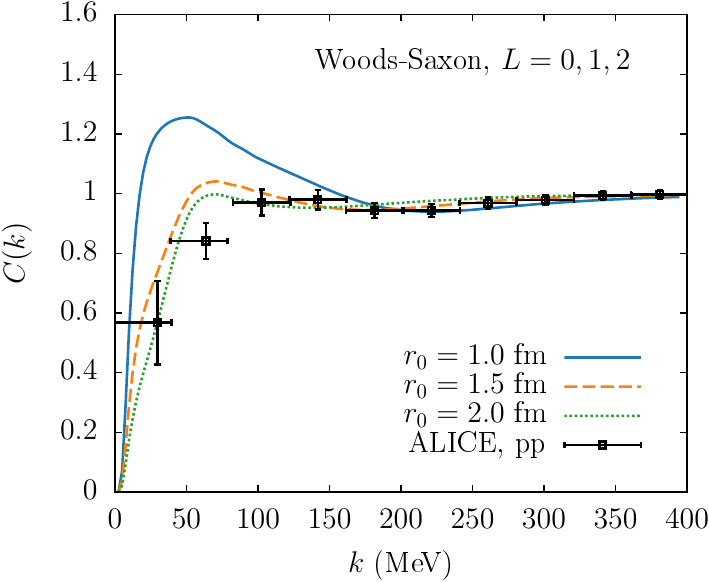}
    \includegraphics[width=0.45\textwidth]{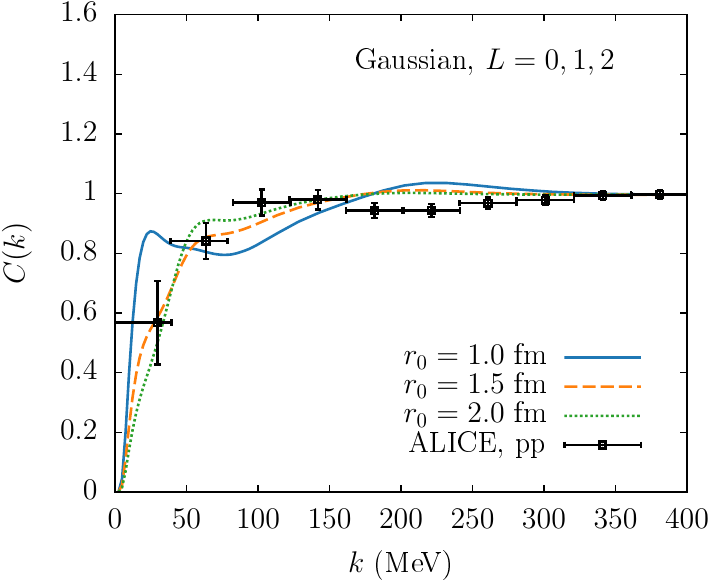}
    \includegraphics[width=0.45\textwidth]{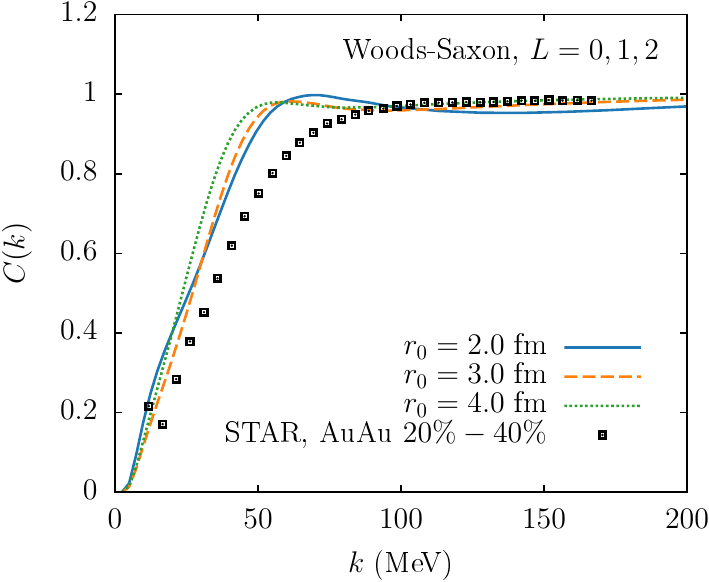}
    \includegraphics[width=0.45\textwidth]{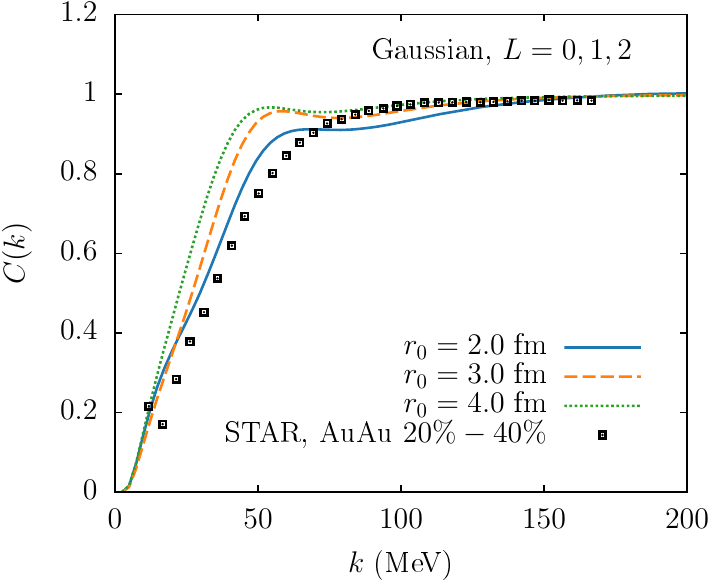}
  \caption{Left column: $pd$ correlation function including $L=0,1,2$ partial waves using the WS potentials, for different values of the Gaussian radius. Right column: Same as left column but using the Gaussian potentials for the $pd$ scattering. We superimposed experimental data from ALICE~\cite{ALICE:2024bhk} and STAR~\cite{STAR:2024lzt} collaborations.}
  \label{fig:partialwaves}
\end{figure}

After discussing the effect of the sequential addition of partial waves, we present the results with $L_{\textrm{max}}=2$ as a function of the potential used (WS versus Gaussian) and the source radius, $r_0$. In  the left column of Fig.~\ref{fig:partialwaves}, we present the results using the WS potentials (with a strongly attractive $^4S_{3/2}$ channel), while in the right column, we show the results using the Gaussian potential (with a repulsive $^4S_{3/2}$ channel). In the top panels, we focus on $p-d$ correlation functions with small source radii $r_0=1,1.5, 2$ fm, relevant for small systems, like those produced in high-multiplicity p+p collisions at the LHC. We superimpose the experimental results from the ALICE collaboration for this system~\cite{ALICE:2023bny}. With a source size  of $r_0=1.0$ fm, we observe a prominent peak in the WS model that cannot describe the data, while the Gaussian potential with the same source radius improves the description significantly. Sources with $r_0=1.5,2$ fm can roughly describe data in both models, but these radii are slightly larger that the effective source radius quoted by the ALICE collaboration in~\cite{ALICE:2023bny}. In any case, our results demonstrate that the two-body description beyond the LL approximation---that is, solving the whole wave function via the Schr\"odinger equation---does a satisfying job.

In the bottom panel we use larger source radii, $r_0=2,3,4$ fm, relevant for larger collision systems. We superimpose data from mid peripheral Au+Au collisions as measured by the STAR Collaboration from BNL~\cite{STAR:2024lzt} (we do not incorporate errors bars, since they are difficult to extract from the figures as they are very small). Our results overpredict the data points, especially in the WS model, possibly due to the extra attraction caused by the $^4S_{3/2}$ bound state. For the Gaussian potential (top, left panel) the agreement improves a bit, thus favouring the repulsive $^4 S_{3/2}$ interaction, but overpredicting the data for a source radius similar to that quoted by STAR collaboration, $r_0\simeq 3$ fm~\cite{STAR:2024lzt}.

We again observe that a two-body potential between the proton and the deuteron provides a reasonable description of the $pd$ correlation function. While a three-body interaction model~\cite{Viviani:2023kxw} results in a more complete and seemingly more precise description of the ALICE data~\cite{ALICE:2023bny}, the simplicity of a two-body problem makes such an approach worthwhile. The claimed failure of the two-body description in Ref.~\cite{ALICE:2023bny} is due to the LL approximation used, and not to a genuine insufficiency of the two-body description, as we have just seen. This statement is further corroborated by the results presented in Appendix~\ref{app:LL}, where we perform an analysis of the correlation function within the LL approximation employing the low-energy parameters that describe the phase shifts of Fig.~\  \ref{fig:phaseshifts}.

\section{Conclusions}

In this work, we have studied the $pd$ system and its femtoscopy under the assumption of a two-body interacting potential. We have calculated the femtoscopy correlation function with source radii relevant for p+p as well as for heavy-ion collisions. To analyze the $p-d$ interaction, we have considered two sets of potentials for the $S=\{1/2,3/2\}$ and $L=\{0,1,2\}$ channels. We have first used Woods-Saxon potentials obtained in Ref.~\cite{Jennings:1985km} using experimental phase shift data from Ref.~\cite{Arvieux:1974fma}. The fitted potentials have a bound state in the $S=1/2,L=0$ channel---the $^3$He---as well as an unidentified bound state in the $=3/2,L=0$ channel. Alternatively, we have fitted Gaussian potentials for the very same channels, but reinterpreting the $^4S_{3/2}$ phase shift in Ref.~\cite{Arvieux:1974fma} as a repulsive interaction, as done in previous works~\cite{VanOers:1967oww,VanOers:1967oww,Eyre:1976dp,Huttel:1983wkj,Kievsky:1996ca}. In this case, we have obtained better fits to the phase shifts with a lesser number of parameters.

The resulting wave functions obtained from the Schr\"odinger equation have been used to predict the $pd$ correlation function for the two models as functions of the radius $r_0$ of a Gaussian source. Overall, we have found a good theoretical description of the experimental data in both p+p and Au+Au collisions using a two-body dynamics. While a three-body problem like the one proposed in Refs.~\cite{Mrowczynski:2019yrr,Viviani:2023kxw} might result on a better approximation to the problem, we find a reasonably good agreement with the data by a much simpler method. 
In previous works~\cite{Viviani:2023kxw,ALICE:2023bny}, the two-body description was claimed to completely fail for the description of the experimental data. However, the correlation function was calculated not by using the complete wave function, but applying the LL approximation~\cite{Lednicky:2005tb}, which takes into account only the asymptotic form of the wave function (in general, only valid for large sources) as we review in Appendix~\ref{app:LL}. We have shown here that employing a two-body dynamics only, but going beyond this approximation, leads to a reasonable description of the data. These results should then be considered as a realistic two-body benchmark that can of course be improved by sophisticated three-body treatment as that of Ref.~\cite{Viviani:2023kxw,ALICE:2024bhk}.

\section{Note added}
After the completion of this work and during the writing of the manuscript, the preprint~\cite{Rzesa:2024oqp} appeared. The motivation and the methodology of this paper coincides with the one used here, and the results are entirely compatible with ours. We fully agree with the conclusions of Ref.~\cite{Rzesa:2024oqp} that the Lednick{\'y}-Lyuboshitz approximation might not be reliable for small sources and, in particular, for correlation functions involving the deuteron.

\begin{figure}[ht]
  \centering
  \includegraphics[width=0.45\textwidth]{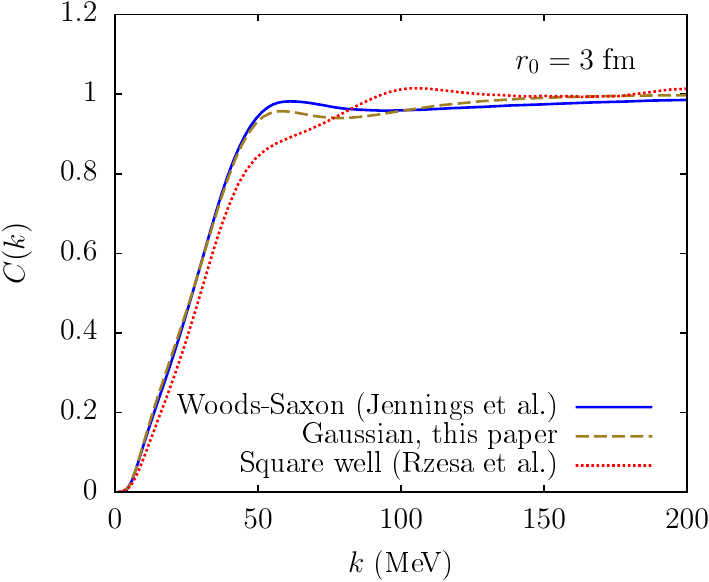}
  \caption{Comparison of the $pd$ correlation function from the solution of the 2-body Schr\"odinger equation with different potentials, the Woods-Saxon potential from Ref.~\cite{Jennings:1985km}, the Gaussian potential parametrized in this work, and a combination of squares wells from the recent Ref.~\cite{Rzesa:2024oqp}. }
  \label{fig:compPratt}
\end{figure}

The potential used in Ref.~\cite{Rzesa:2024oqp} is a combination of square wells, while here we have used Woods-Saxon and Gaussian profiles. For comparison, we present in Fig.~\ref{fig:compPratt} the $pd$ correlation function with a source radius of $r_0=3$ fm taken from Ref.~\cite{Rzesa:2024oqp} together with our calculations with the WS and Gaussian potentials. All calculations are taken up to $L=2$. We observe small differences due to the particular form of the potentials that provide a different behavior of the wave functions for $r \lesssim 3$ fm.

\begin{acknowledgments}

We thank Scott Pratt for an enlightening discussion  after the appearance of Ref.~\cite{Rzesa:2024oqp} on the important role of the $^4 S_{3/2}$ channel in the $pd$ scattering.

This work has been supported by the project number CEX2019-000918-M (Unidad de Excelencia ``Mar\'ia de Maeztu''), PID2020-118758GB-I00 and PID2023-147112NB-C21, financed by the Spanish MCIN/ AEI/10.13039/501100011033/.
JMT-R also thanks the Deutsche Forschungsgemeinschaft (DFG) through the CRC-TR 211 “Strong- interaction matter under extreme conditions” for support.

\end{acknowledgments}

\appendix

\newpage 
\section{Lednick{\'y}-Lyuboshitz approximation~\label{app:LL}}

In the Lednick{\'y}-Lyuboshitz approximation (LL)~\cite{Gmitro:1986ay}, the pair wave function $\psi(\bm{r};\bm{k})$ going into the Koonin-Pratt formula is taken in its asymptotic form, even for small values of $\bm{r}$. In the nonrelativistic case and without the electromagnetic interaction, this amounts to using~\cite{joachain1975quantum},
\begin{equation} \psi (\bm{r};\bm{k} ) = e^{i \bm{k} \cdot \bm{z}} + f (k) \frac{e^{ikr}}{r} \ , \end{equation}
where $f (k) \equiv f(\bm{k} \rightarrow \bm{k'})|_{|\bm{k}|=|\bm{k}'|}$ is the scattering amplitude. This quantity contains all the information on the strong interaction, and in the standard use of the LL approximation, only the $L=0$ partial wave is taken. See, for example, the application to the proton-deuteron case by the ALICE collaboration in Ref.~\cite{ALICE:2023bny}. In this particular system, the Coulomb interaction should also be considered and the total wave function $\Phi(r,z;k)$ needs to be modified according to the Coulomb case~\cite{joachain1975quantum,Torres-Rincon:2023qll},
\begin{equation} \Phi(r,z;k) = \Phi^\textrm{C}(r,z;k)+ e^{-\pi \gamma/2} \Gamma(1+ i\gamma)  G(\gamma)^{1/2} f_C (k) \frac{i F_0 (\gamma;kr)- G_0 (\gamma;kr)}{r}\ , 
\label{eq:A2}
\end{equation}
where $\Phi^C(r,z;k)$ is the total Coulomb wave function given in Eq.~\eqref{eq:completeCoulomb}, $\gamma$ is the Sommerfeld factor $\gamma=Z_1 Z_2 \mu \alpha/k$, $G(\gamma)$ is the Gamow factor,
\be G(\gamma) = \frac{2\pi \gamma}{e^{2\pi \gamma}-1} \ , \ee
$F_0$ and $G_0$ are the regular and irregular $L=0$ Coulomb wave functions, respectively, and $f_C(k)$ is the scattering amplitude in the presence of the electromagnetic force. In Ref.~\cite{Gmitro:1986ay} the low-energy expansion of $f_{\rm{C}} (k)$ is also used,
\begin{equation} f_{\rm{C}}^{-1} (k)= - \frac{1}{a_0} + \frac12 d_0 k^2 - 2 k \gamma h( \gamma^{-1})- i k G(\gamma)   \ , \label{eq:fC} \end{equation}
where 
\be h(x^{-1})= - \log(|x|)+ \frac12 \psi(1-ix) + \frac12 \psi(1+ix) \ , \ee
with $\psi(x)$ the digamma function. 

Because the presence of the irregular function $G_0 (\gamma, kr)$ in Eq.~(\ref{eq:A2})  forbids the extension of the asymptotic form to small values of $kr$, in Ref.~\cite{Gmitro:1986ay} a further approximation is made for $\gamma k r \ll 1$ in which ${_1}F_{1}(-i\gamma;1;ik(r-z)) \simeq 1$, $F_0(\gamma;kr) \simeq G(\gamma)^{1/2} \sin(kr)$ and $G_0(\gamma;kr) \simeq - G(\gamma)^{-1/2} \cos(kr)$.

With these modifications, and using the Gaussian source in Eq.~\eqref{eq:source} we have obtained an analytic formula for the LL approximation with Coulomb interaction in terms of the strong scattering length and scattering range,
\begin{equation} C_{\rm{LL}} (k)= G(\gamma) \left\{ 1 + \frac{2 \textrm{Re} f_{\rm{C}}} {\sqrt{\pi} r_0} F_1 (2qr_0) 
 - \textrm{Im} f_{\textrm{C}} \frac{G(\gamma)}{r_0 } F_2(2qr_0)
+ \frac{|f_C|^2}{2r_0^2} \left[ 1+ \frac12 [G(\gamma)^2 -1]
(1 -  e^{-4q^2r_0^2})  \right] \right\} \ ,\label{eq:CkLL} \end{equation}
with $F_1(x)=x^{-1}\exp(-x^2)\int_0^x \exp(y^2) dy$ and $F_2(x)=(1-\exp(-x^2))/x$.

To apply Eq.~\eqref{eq:CkLL} to our results, we first extract the low-energy parameters ($a_0^{1/2},d_0^{1/2},a_0^{3/2},d_0^{3/2}$) by adjusting the following expression
\be k \cot \delta^J_0(k)= - \frac{1}{a^J_0} + \frac12 d_0^J k^2 -2 q \gamma h(\gamma^{-1}) \ , \label{eq:effrange} \ee
to the phase shifts shown in Fig.~\ref{fig:phaseshifts}, namely the original Arvieux data~\cite{Arvieux:1974fma} and those obtained using the Woods-Saxon and Gaussian potentials (which we recall were precisely fitted to the former data).
For further reference, we note that we use the same sign convention for the scattering length as in Ref.~\cite{Arvieux:1974fma}, consistent also to the sign in the expression of Eq.~\eqref{eq:fC} (which is the opposite convention as in our Ref.~\cite{Torres-Rincon:2023qll}).

We summarize the resulting low-energy parameters in Table~\ref{tab:lowene}. 

\begin{table}[ht!]
\centering
\vspace*{2mm}
\begin{tabular}{|c|c|c|c|c|}
\hline
&   \multicolumn{2}{c|}{$L=0,S=1/2$} & \multicolumn{2}{c|}{$L=0,S=3/2$}  \\
\hline
 & $a_0$ (fm) & $d_0$ (fm)&  $a_0$ (fm) & $d_0$ (fm)  \\
\hline
Arvieux & $(2.72 \pm 0.08)$ & $(2.31 \pm 0.10)$ & $(11.93 \pm 0.25)$ & $(2.50 \pm 0.04)$\\
Woods-Saxon & $3.68$ & $1.44$ & $11.94$ & $2.47$  \\
Gaussian & $4.07$ & $1.71$ & $10.95$ & $2.52$  \\
\hline
\end{tabular}
\caption{Extracted low-energy parameters in the data and models used in this work by fitting the $pd$ phase shifts according to Eq.~\eqref{eq:effrange}. ``Arvieux'' refers to the experimental data given in Ref.~\cite{Arvieux:1974fma}, and ``Woods-Saxon'' and ``Gaussian'' to the phase shifts obtained with these two potentials. \label{tab:lowene}}
\end{table}

The straight fit to Arvieux's data---within the same momentum range as indicated in Ref.~\cite{Arvieux:1974fma}---is fully consistent with the numbers quoted in that reference. The fits to the phase shifts generated by the Woods-Saxon and Gaussian potentials are useful to quantify the systematic uncertainty due to the choice of the microscopic model. We notice that the stronger dispersion of the extracted low-energy parameters appears in the $S=1/2$ channel, while those for $S=3/2$ are pretty consistent among each other.

\begin{figure}[ht]
  \centering
 \includegraphics[width=0.65\textwidth]{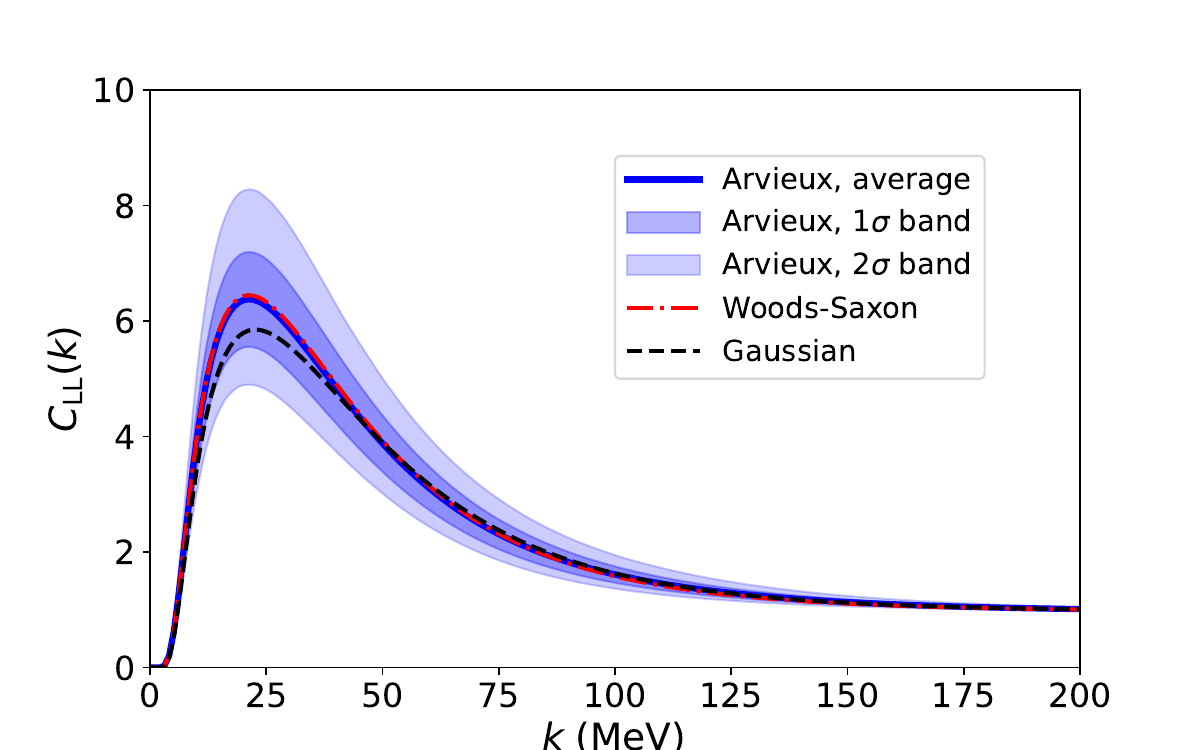}
  \caption{Proton-deuteron correlation function in the LL approximation using the low-energy parameters of the data and models shown in Table~\ref{tab:lowene}. See details in the main text.
  \label{fig:LL}}
\end{figure}

We finally implement these low-energy scattering data into the LL formula in Eq.~\eqref{eq:CkLL} and generate the $pd$ correlation function in this approximation. We present our results in Fig.~\ref{fig:LL}.

For the Arvieux case, we plot the correlation function in the LL approximation using the average values of the low-energy parameters (blue solid line) and we consider their uncertainties, together with that of the Gaussian radius $r_0=(1.08 \pm 0.06)$ fm taken from Ref.~\cite{ALICE:2023bny}, to generate error bands of 1$\sigma$ and 2$\sigma$ confidence levels around the average result. For the Woods-Saxon (red, dotted-dashed line) and Gaussian (black dashed line) potentials, we observe that the two curves lie within the $1\sigma$  error band of the Arvieux case.
Our results are fully consistent with the LL calculation shown in the ALICE paper~\cite{ALICE:2023bny}. We can therefore conclude that the correlation function in the Lednick{\'y}-Lyuboshitz approximation gives an extremely different prediction of the correlation function from the one obtained in the present paper, using the complete two-body wave function instead of the asymptotic one, and from the experimental result of Ref.~\cite{ALICE:2023bny}.

\bibliography{references}

\end{document}